\documentclass[aps,prl,superscriptaddress,reprint,amsmath,amssymb]{revtex4-2}

\usepackage{define}

\newcommand{\Hub}{\text{Hub}}
\newcommand{\sHub}{\text{$\sigma$-Hub}}

\newcommand{\prlsection}[1]{\textit{#1}.---}

\begin{document}

\title{Hubbard Model on Triangular Lattice: Role of Charge Fluctuations }

\author{Ji-Si Xu}
\affiliation{Institute for Advanced Study, Tsinghua University, Beijing 100084, China}

\author{Zheng Zhu}
\email{zhuzheng@ucas.ac.cn}
\affiliation{Kavli Institute for Theoretical Sciences, University of Chinese Academy of Sciences, Beijing 100190, China}
\affiliation{CAS Center for Excellence in Topological Quantum Computation, University of Chinese Academy of Sciences, Beijing, 100190, China}

\author{Kai Wu}
\affiliation{Manifold Creative Global, 100 Lorong 23 Geylang, 388398, Singapore} 

\author{Zheng-Yu Weng}
\email{weng@mail.tsinghua.edu.cn}
\affiliation{Institute for Advanced Study, Tsinghua University, Beijing 100084, China}

\date{June 19, 2023}

\begin{abstract}

A chiral spin liquid (CSL) phase has been recently reported in the Hubbard model on a triangular lattice at half-filling. It emerges in an intermediate coupling regime, which is sandwiched between a 120\degree\ antiferromagnetic (AFM) phase and a metallic phase as a function of on-site repulsion $U$. 
In this work, we examine how the charge fluctuation may cause the CSL and complex phase diagram. We first identify an exact sign structure of the model at arbitrary temperature, sample size, and doping, which is reduced  from the original Fermi exchange signs of the electrons by finite $U$. In particular, the spin and charge degrees of freedom are generally entangled via the phase-string in the sign structure. 
By precisely switching off such a phase-string, the CSL phase is shown to disappear, with only the 120\degree\ AFM order remaining down to weak $U$ by a density matrix renormalization group calculation. Here the charge fluctuation is effectively decoupled from the AFM background, which is a 120\degree\ AFM order due to a geometric Berry phase in the sign structure of the triangular lattice. General implications for the Mott physics will also be discussed.

\end{abstract}

\maketitle

\prlsection{Introduction}%
The quantum spin liquid (QSL) has been a subject with growing interest since the concept was first introduced by Anderson in the 1970s \cite{Anderson1973Resonating,Balents2010Spin,Savary2017Quantum,Zhou2017Quantum,Knolle2019Field,Broholm2020Quantum}. 
The antiferromagnetic (AFM) spin system on triangular lattice was once believed to be a promising candidate to realize a QSL state due to its strong geometric frustration. But the numerical studies have revealed that in the Heisenberg model, the ground state is actually 120\degree\ AFM long-range ordered \cite{Capriotti1999LongRange,White2007Ne}. 
Experimentally some triangular lattice materials like $\text{Ba}_3\text{Co}\text{Sb}_2\text{O}_9$ \cite{Shirata2012Experimental,Susuki2013Magnetization,Koutroulakis2015Quantum,Quirion2015Magnetic,Ma2016Static,Ito2017Structure} clearly exhibit a 120\degree\ AFM order. 
On the other hand, several candidate materials with triangular lattice, such as the organic Mott insulators $\kappa\text{-}(\text{BEDT-TTF})_{2}\text{Cu}_2(\text{CN})_3$ \cite{Shimizu2003Spin,Kurosaki2005Mott,Yamashita2008Thermodynamic,Yamashita2009Thermaltransport} and $\text{Et}\text{Me}_3\text{Sb}[\text{Pd}(\text{dmit})_2]_2$ \cite{Itou2008Quantum,Yamashita2010Highly,Yamashita2011Gapless}, have been considered potentially as the QSL systems. 

In contrast to the Heisenberg model with spin local moment, a QSL phase in a Hubbard model at half-filling with an intermediate strength of $U$ has been recently reported \cite{Yoshioka2009Quantum,Laubach2015Phase,Misumi2017Mott,Shirakawa2017Groundstate,Szasz2020Chiral,Szasz2021Phase,Chen2022Quantum}.
Such a QSL phase is located between the metallic regime at small $U$ and the 120\degree\ AFM phase at large-$U$, where the latter is continuously connected to the Heisenberg model in the large-$U$ limit. 
In particular, this QSL has been further identified as a chiral spin liquid (CSL) \cite{Szasz2020Chiral,Szasz2021Phase}.

Whereas the geometric frustration in a Heisenberg model alone is not enough to drive the system into a quantum disordered phase, the QSL in a Hubbard model may be attributed to the fact that an increasing charge fluctuation plays an important role. 
Among the numerous theories for QSL, the interplay between the spin and the charge degrees of freedom has been discussed \cite{Lee2005U,Ng2007PowerLaw,Senthil2008Theory,Podolsky2009Mott}.
But most theories have assumed \textit{a priori} the existence of QSL, with the focus on the Mott transition to a metallic phase \cite{Senthil2008Theory,Podolsky2009Mott}. 
Here a high-order spin interaction, which emerges from the charge fluctuation \cite{MacDonald1988FractU}, may stabilize the QSL state. 
The numerical results \cite{Cookmeyer2021FourSpin} show that a Heisenberg model with four spin term or other additional terms can give rise to a rich phase diagram, including the QSL phase in the strong coupling regime. 
But how exactly the charge fluctuations influence the spin order and \emph{vice versa}, including the Mott transition at smaller $U/t$, still remains to be understood. 

In this paper, we investigate the mechanism of how the charge fluctuation frustrates the 120\degree\ AFM order as $U$ reducing from the strong-coupling limit. 
We first analytically identify the precise sign structure of the triangular lattice Hubbard model at an arbitrary $U$, temperature, and doping concentration based on the partition function.
In particular, we show that the charge and spin degrees of freedom are intrinsically entangled by a novel sign structure known as the phase-string, besides the conventional fermions statistics between the dopants and some geometric Berry phase associated with the triangular lattice. The same kind of phase-string factor has been previously identified on the square lattice for the Hubbard model without the additional frustration of the geometric phase \cite{Zhang2014Sign}. Its large-$U$ version has also been previously found in the $t$-$J$ model for both square lattice \cite{Sheng1996Phase, Wu2008Sign} and triangular lattice \cite{Wu2011Statistical}. 

Then by ``switching off'' the phase-string (without changing the other phase factors and path-dependent weight) in a density matrix renormalization group (DMRG) calculation, we show that the CSL disappears in the phase diagram, with only the AFM state persisting all the way down to the weak $U$. 
It thus clearly indicates that the charge fluctuations are mediated by the phase-string sign structure to twist the AFM into a QSL and eventually the collapse of the charge gap in the triangular Hubbard model at half-filling. Namely, the Mott physics is generally dictated by the novel  sign structure of the Hubbard model. 

\begin{figure}[tbp]
    \centering
    \includegraphics[width=3.4in]{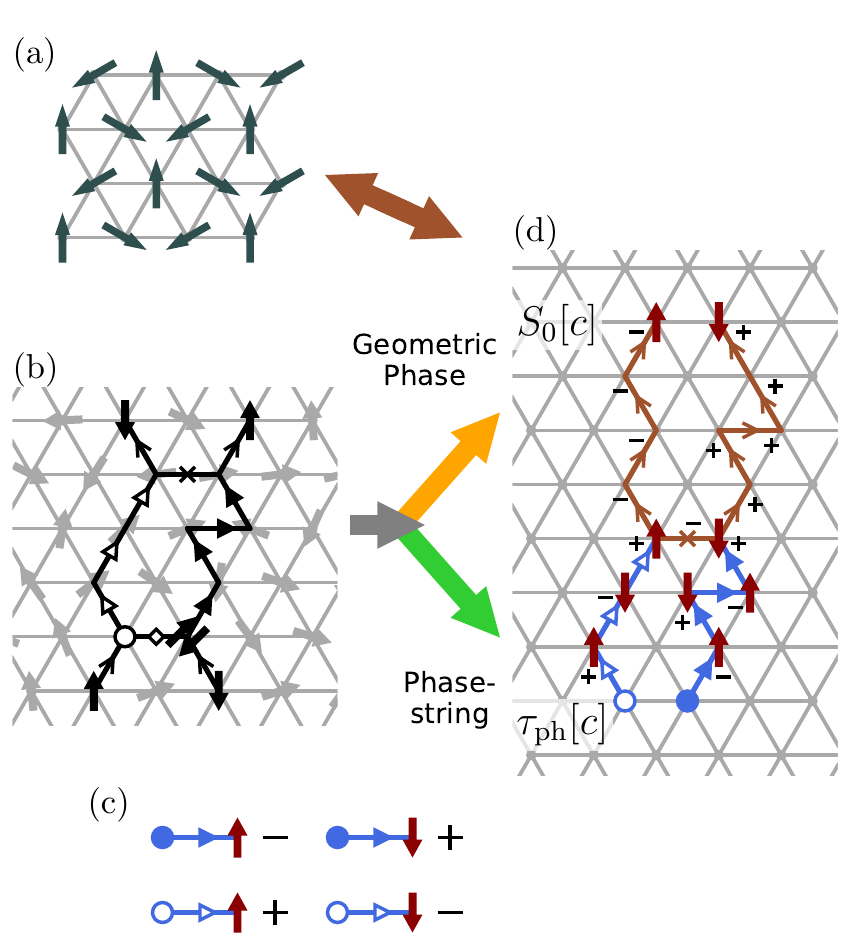}
    \caption{\label{fig:summary} 
    The illustration of the sign structure of the triangular Hubbard model as given in Eq. (\ref{eq:SPStri}).
    (a) A typical spin configuration of 120\degree\ AFM order at large-$U$; 
    (b) Charge fluctuations as spontaneous creation (at link marked by an open diamond) and annihilation (at link marked by the cross) of the holon (open circle) and doublon at half-filling; 
    (c) Elementary processes of the hopping of the chargon (doublon, filled circle and holon, open circle) and the associated signs, $\pm$, depending on the spin swapped with the chargon. 
    They contribute to the phase-string $\tau_{\mathrm{ph}}$ in Eq. (\ref{eq:tau}) for the chargon hoppings as illustrated in (d); 
    Each up-spin hopping on the lattice will acquire an additional geometric ($-$) sign [cf. (d)], which gives rise to a geometric phase $S_0$ in Eq. (\ref{eq:S0}); 
    Finally, it is noted that the chargons will contribute to an additional minus sign in Eq. (\ref{eq:SPStri}) each time two identical holons or doublons are exchanged as if they are fermions (not shown here in the figure).}
\end{figure}

\prlsection{Sign Structure of the Hubbard Model}%
In this work, we shall study the Hubbard model, 
\begin{equation}
H_\Hub = H_t+ U \sum_{i} n_{i\SU}n_{i\SD}       \label{eq:Hubbard}
\end{equation}
where $H_t\equiv - t \sum_{\nn{ij},\sigma} \cd_{i\sigma}\co_{j\sigma}+h.c.$ denotes the nearest neighbor (NN) hopping on a triangular lattice.  

We shall first identify an exact sign structure in the partition function as follows
\begin{gather}
Z_\Hub \equiv \tr(\ee^{- \beta H_\Hub}) = \sum_{c} S[c] W[c]
\label{eq:ZtoSW}
\end{gather}
where $W[c]\ge 0$ is a positive weight, while $S[c]$ with $|S[c]|=1$ denotes the sign structure for any closed loop $c$ of the spin and charge coordinates in the full Hilbert space. 
The proof is based on the high-temperature ($T=1/\beta$) series expansion of the partition function to all orders:
\begin{gather}\label{eq:Zhub}
Z_\Hub = 
\sum_{n=0}^{\infty} \frac{\beta^n}{n!} 
    \sum_{\set{\alpha_i}_{i=1}^n} 
        \prod_{i} \bra{\alpha_{i+1}} (- H_\Hub)\ket{\alpha_i} 
\end{gather}
where $\alpha$ is the label of a complete set of basis composed of the spin(on) (in the $\mathrm{S}^z$-quantization) and chargon coordinates at single occupied and empty/double occupied sites (see the Supplemental Material in Ref. \onlinecite{SM}). 
We can view $c = \set{\alpha_i}_{i=1}^n$ as a closed loop in the coordinate space with $\ket{\alpha_1}= \ket{\alpha_{n+1}}$.

Here $S[c]$ collects all the signs of the matrix elements $\bra{\alpha_{i+1}} (- H_\Hub)\ket{\alpha_i}$ to give rise to \cite {SM}
\begin{gather}
S[c] \equiv S_0[c]  \times (-1)^{N_{\mathrm{ex}}^{\mathrm{ch}}[c]} \times \tau_{\mathrm{ph}}[c]
\label{eq:SPStri}
\end{gather}
where
\begin{gather}
S_0[c] \equiv (-1)^{N^\SU[c]}~.
\label{eq:S0}
\end{gather}
Here $N^\SU[c]$ denotes the total steps of hopping of $\SU$-spinons in a closed loop $c$; $(-1)^{N_{\mathrm{ex}}^{\mathrm{ch}}[c]}$ counts the fermion signs pending on $N_{\mathrm{ex}}^{\mathrm{ch}}[c]$ as the total exchange number between the holons and between the doublons as the identical particles. The third factor in Eq. (\ref{eq:SPStri}) is the most exotic which indicates a \emph{mutual statistics} between the chargons and spinons   
\begin{gather}
\tau_{\mathrm{ph}}[c] \equiv (-1)^{N_{\SD}^{h}[c]} \times (-1)^{N_{\SU}^{d}[c]}
\label{eq:tau}
\end{gather}
in which $N_{\SD}^{h}[c]$ and $N_{\SU}^d[c]$ count the total number of swaps between the chargons and spinons, i.e., a holon and a $\SD$-spinon, and a doublon and an $\SU$-spinon, respectively, in the closed loop $c$. 
The basic processes within a closed loop $c$ and the associated sign structure are figuratively illustrated in Fig. \ref{fig:summary}. 

It is noted that previously a similar sign structure has been exactly identified \cite{Zhang2014Sign} for the 2D Hubbard model on a square lattice, in which $S_0[c] = 1$ in Eq. (\ref{eq:SPStri}) without the geometric frustration. 
Here for the triangular lattice, $S_0[c] $ is nontrivial which depends on the parity of the total triangular units enclosed within the closed paths of the $\SU$-spinons. 
In the large-$U$, i.e., the Heisenberg limit at half-filling, it will be responsible for driving the system into the 120\degree\ AFM order (see below). 
The fermion statistical sign factor associated with the chargons is also conventional, which is similar to doping a semiconductor. 
Inside the sign structure in Eq. (\ref{eq:SPStri}), the sign factor $\tau_{\mathrm{ph}}[c]$ introduces a novel long-range mutual entanglement between the spin and charge degrees of freedom, which is known as the phase-string whose nontrivial effect has been previously studied in the doped cases on a square lattice in the large-$U$ limit\cite{Weng1997Phase,Weng2011Superconducting}. 

To examine its unique effect, one may exactly switch off $\tau_{\mathrm{ph}}[c]$ in Eq. (\ref{eq:SPStri}), with the partition function reducing to
\begin{gather}
Z_\sHub \equiv \tr(\ee^{- \beta H_\sHub}) \equiv \sum_{c} S_0[c] (-1)^{N_{\mathrm{ex}}^{\mathrm{ch}}[c]} W[c]
\label{eq:Zsigma}
\end{gather}
with the same weight $W[c]$ \cite{SM}. It is straightforward to show that the corresponding Hamiltonian is modified as \cite{SM}
$H_\sHub = H_{\sigma t}+ U \sum_{i} n_{i\SU}n_{i\SD}$
where 
\begin{gather}
H_{\sigma t} 
\equiv - t \sum_{\langle ij \rangle,\sigma} 
    \cd_{i\sigma}\co_{j\sigma}[\sigma \hat{P}^{T}_{ij} + (1-\hat{P}^{T}_{ij})] 
    + \HC  
\label{eq:Hst}
\end{gather}
with $\hat{P}^{T}_{ij}$ a projection operator to enforce a single chargon (holon or doublon) at the NN bond $ij$, whose hopping involves an exchanging with a spinon [cf. Fig. \ref{fig:summary} (c)]. 
By contrast, the projection $(1-\hat{P}^{T}_{ij})$ involves a simultaneous creation or annihilation of a pair of holon-doublon at $ij$. 

Therefore, the sole distinction between the Hubbard and $\sigma$-Hubbard models lies in the presence and absence of the phase-string factor of Eq. (\ref{eq:tau}) in their respective sign structures [cf. Eqs. (\ref{eq:ZtoSW}) and (\ref{eq:Zsigma})]. Physically the phase-string of Eq. (\ref{eq:tau}) will dynamically entangle the charge and spin degrees of freedom in the Hubbard model, which otherwise may behave independently of each other in a more conventional manner in the $\sigma$-Hubbard model (see below).

\begin{figure}[tp]
    \centering
    \includegraphics[width=3.2in]{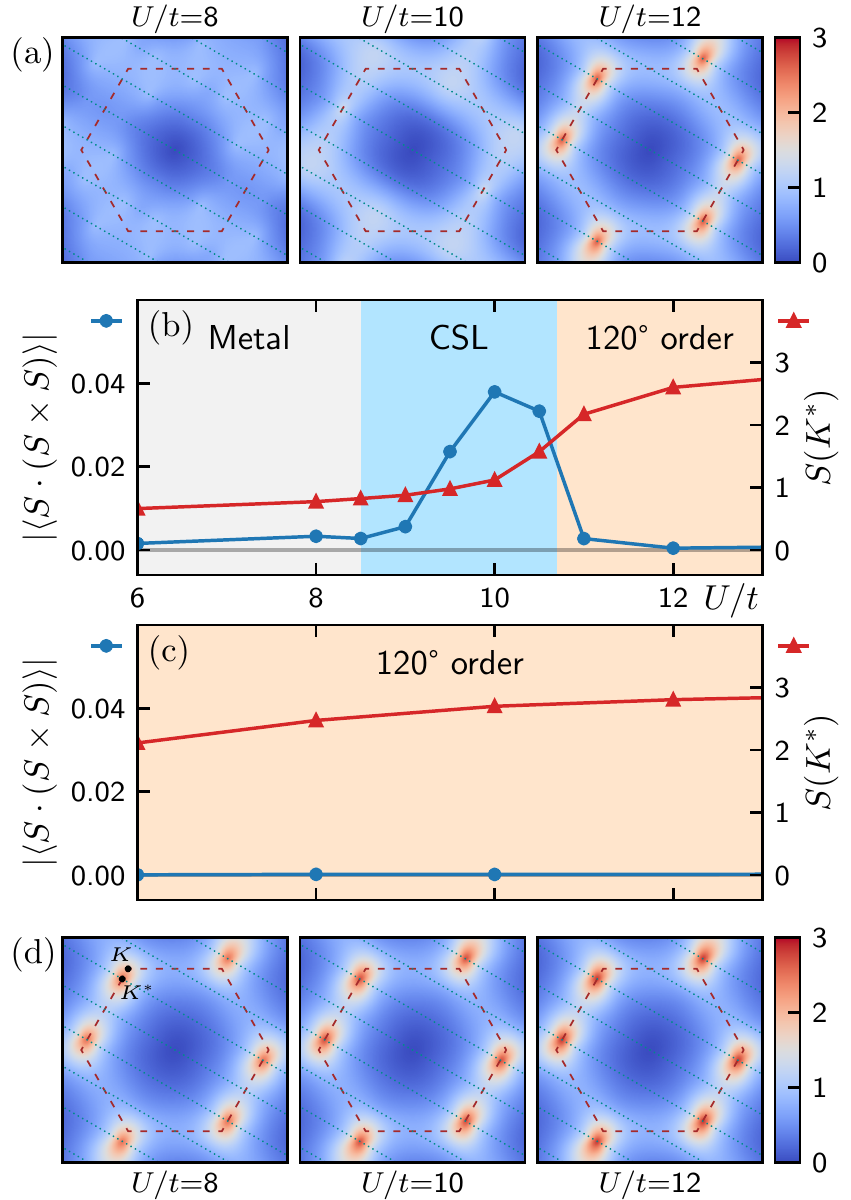}
    \caption{\label{fig:spin} 
    The spin characterizations of the phase diagram in the triangular Hubbard model and $\sigma$-Hubbard model by DMRG calculation. 
    (a) and (d): The momentum distributions of the spin structure factor $S({\mathbf{q}})$ at $U/t=8,10,18$, respectively. The three distinct phases of the Hubbard model in (a) reduce to a single phase in the $\sigma$-Hubbard model [(d)]; 
    (b) and (c): The chiral order parameter $|\ave{\mathbf{S}_i \cdot (\mathbf{S}_j \times \mathbf{S}_k)}|$ (blue) and the spin structure factor at $K^*$ (red), for the Hubbard model [(b)] and the $\sigma$-Hubbard model [(c)], respectively.}
\end{figure}

\prlsection{Phase diagram at half-filling: DMRG results}%
To precisely characterize the distinction between the Hubbard model and $\sigma$-Hubbard models at half-filling, we employ the DMRG algorithm to examine the ground state properties. The triangular lattice is spanned by the primitive vectors  $\mathbf{e_x}=(1,0), \mathbf{e_y}=(1/2,\sqrt{3}/2)$ and wrapped on cylinders with circumference of 4. Depending on the nature of distinct phases, the bond dimension $\mathcal{D}$ is pushed up to $\mathcal{D}=24000$ to secure the convergence.  

\begin{figure}[tp]
    \centering
    \includegraphics[width=3.4in]{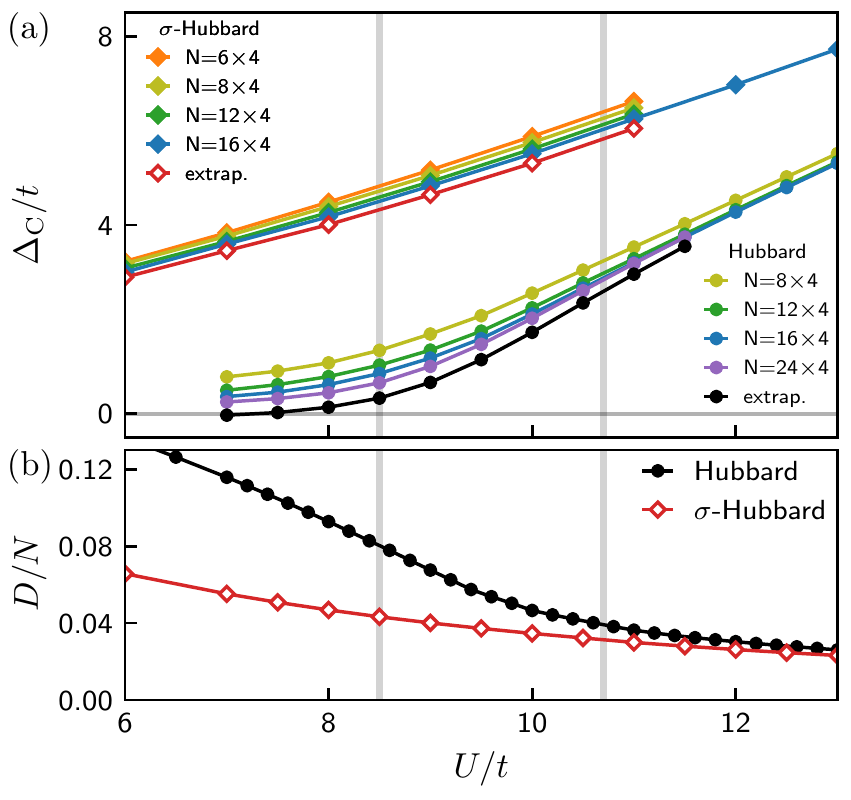}
    \includegraphics[width=3.4in]{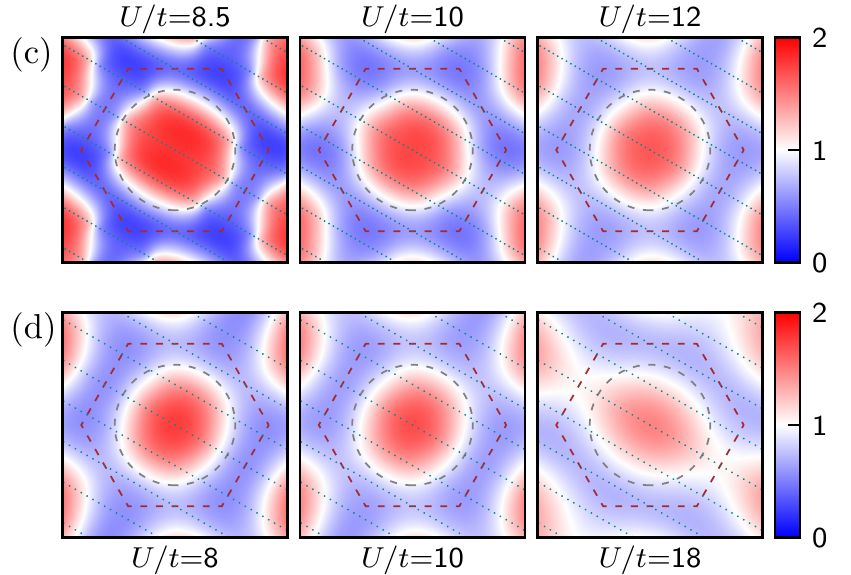}
    \caption{\label{fig:charge} 
    The charge characterizations of the Hubbard and $\sigma$-Hubbard models by DMRG calculation. (a): The charge gap; (b): The average double occupancy number per site $D/N$. The vertical lines indicate the two phase transition points of the Hubbard model; (c) Electron momentum distribution $n_k$ for the Hubbard model; (d) $n_k$ for the $\sigma$-Hubbard model.} 
\end{figure}

In Fig.~\ref{fig:spin}, the characteristics of the spin degrees of
freedom are shown in the intermediate $U/t$ regime.
The upper panels, (a) and (b), present the results for the Hubbard model, in which three typical phases are shown. 
To identify the 120\degree\ AFM order, we compute the spin structure factor $S(\mathbf{q})$, which is defined as
 $S(\mathbf{q}) =1/N\sum_{ij}\langle \mathbf{S}_{i}\cdot\mathbf{S}_{j}\rangle  e^{i\mathbf{q}\cdot(\mathbf{r}_{i} - \mathbf{r}_{j})}$.
As shown in Fig.~\ref{fig:spin} (a),  $S(\mathbf{q})$ is peaked at $\mathbf{q}= \mathbf{K^*}$, characterizing the 120\degree\ AFM order at large-$U$ side, 
here $\mathbf{K^*}$ is the closest allowed momentum to $\mathbf{K}$ as the characteristic momentum of the 120\degree\ AFM order.
Then a CSL order as characterized by the order parameter $\lvert\langle \mathbf{S}_i \cdot (\mathbf{S}_j \times \mathbf{S}_k)\rangle\rvert$ sets
in over an intermediate regime approximately between $8.5 < U/t < 10.7$ [see Fig. \ref{fig:spin}(b)], where the peaked spin structure factor $S(\mathbf{K^*})$ gets diminished.
As $U/t$ continues to decrease further, the CSL order eventually vanishes and the system enters a metallic phase at $U/t \approx 8.5$. 
Such an insulator-to-metal transition can also be identified by the close of the charge
gap $\Delta_{c}=\frac{1}{2}[ E_{0}(N_{\uparrow}+1, N_{\downarrow}+1) + E_{0}(N_{\uparrow}-1, N_{\downarrow}-1)- 2E_{0}(N_{\uparrow}, N_{\downarrow})]$,
as shown in Fig. \ref{fig:charge}(a), here $E_{0}(N_{\uparrow}, N_{\downarrow})$ denotes the ground-state energy of a system with $N_{\uparrow}$ spin-up electrons and $N_{\downarrow}$ spin-down electrons, and $N_{\uparrow}=N_{\downarrow}=N/2$ at half filling.
We remark that these results are consistent with the previous study \cite{Szasz2020Chiral}.

By contrast, the corresponding DMRG results for the $\sigma$-Hubbard model are also presented in the lower panels of Fig. ~\ref{fig:spin} in (d) and (c), respectively. 
As one can see, the phase diagram is totally changed from that of the Hubbard model: 
in the whole parameter regime of $U/t$ that we inspect, the 120\degree\ AFM order always remains dominant as the sole stable phase with no more CSL phase. 
Here the charge gap remains finite (cf. Fig.~\ref{fig:charge}(a)) and is persistent down to $U/t\approx2$. 
Furthermore, in Fig.~\ref{fig:charge}(b) the number of double-occupancy per
site, $D/N$, exhibits a fast increase in the Hubbard model in the CSL region from the larger $U$. 
For the $\sigma$-Hubbard model, however, $D/N$ evolves much more smoothly and flatly over the whole region. 
It indicates that the charge fluctuations are well decoupled from the 120\degree\ AFM spin order in the latter. One does encounter a phase transition from strong $U$ until $U/t$ reducing to the order of one, in sharp contrast to the three phases identified at $U/t>8$ in the Hubbard case as illustrated in Fig. ~\ref{fig:charge}.  

\prlsection{Discussion}%
The conventional Fermi sign structure of the electrons can be exactly re-organized in terms of the spin(on) and chargon (holon and doublon) coordinates. Such a new sign structure in Eq. (\ref{eq:SPStri}) becomes useful with turning on the on-site Coulomb repulsion $U$ to differentiate the charge and spin degrees of freedom. 
For example, in the large-$U$ limit with the chargon excitations monotonically suppressed at half-filling, the residual sign structure is essentially bosonic with only a Berry phase left [Eq. (\ref{eq:S0})], which causes a geometric frustration for a triangular lattice to lead to the 120\degree\ AFM order for the spins. With reducing $U$ to the intermediate regime ($\sim 8.5<U/t<10.7$), the AFM order is driven into the CSL phase, and eventually into a metallic phase at smaller $U/t\lesssim8.5$ with closing the Mott gap. 
The underlying mechanism is revealed as due to the charge frustration mediated through the phase-string component of Eq. (\ref{eq:SPStri}). Indeed, if one exactly turns off the phase-string to result in the $\sigma$-Hubbard model [cf. Eq. (\ref{eq:Zsigma})], the DMRG calculation shows that the 120\degree\ AFM order will persist all the way down to a much smaller $U/t$ without the CSL or metallic phase transition. It means that the correct route to a QSL is via the charge fluctuations rather than a geometric frustration with the phase-string playing the key role. 

It is important to realize that for both the Hubbard and $\sigma$-Hubbard models, the amplitude $W[c]$ for each path $c$ in the partition functions remains the same. It depends on the amplitudes of $t$, $U$, and temperature, and is expected to be a smooth functional of the path $c$, which is in sharp contrast to the phase-string sign structure [Eq. (\ref{eq:tau})]. The latter is singular as its sign changes with merely a spin-flip in the total spins exchanging with a chargon for any path $c$. As the sole distinction between the two models, the phase-string strongly influences both the spin and charge sectors of the Hubbard model by a quantum interference effect under the summation of all the closed paths in an intermediately strong $U$. 
Similar to the square lattice case \cite{Wu2008Sign,Zhang2014Sign}, by a duality transformation, one may exactly map the phase-string effect into a topological gauge structure \cite{Weng1997Phase,Weng2011Superconducting,Zhang2014Sign} in which the fractionalized spin and charge degrees of freedom are mutually coupled, and this framework has been previously generalized to the triangular lattice $t$-$J$ model at large doping \cite{Wu2011Statistical}. It will be very interesting to see how the CSL and the metallic phase as revealed by DMRG may naturally arise from such a gauge interaction, which will be explored elsewhere by a perturbative approach.    

Furthermore, the sign structure identified here for the triangular Hubbard model is exact at arbitrary $U$, doping, and temperature, as well as sample size and dimensionality. Therefore, a systematic exploration based on the Hubbard and $\sigma$-Hubbard models using the finite-size exact numerical methods may be also very useful to understand such strongly correlated systems at finite doping \cite{Zhu2022Doped,Chen2022Proposal,Song2021Doping,Gannot2020Hubbard,Wietek2022Tunable,Zampronio2022Chiral,Zhang2023Pseudogap,Kadow2022Hole,Gneist2022Competing}. 
Recently a contrasted DMRG study based on the sign structure has provided new insights into the origin of superconducting and charge-density-wave orders at finite doping in the $t$-$t'$-$J$ model on square lattice\cite{Lu2023Sign}. A similar approach for the triangular lattice may also be interesting in the large-$U$ limit at finite doping \cite{Huang2023Quantum,Zhu2023Superconductivity,Chen2023Singlet,Schlomer2023Kinetictomagnetic,Jiang2021Superconductivity}, where the phase-string effect\cite{Weng1997Phase,Weng2011Superconducting} associated with the doped holes/electrons in Eq. (\ref{eq:SPStri}) becomes singularly important. 

{\it Acknowledgments.---}
We acknowledge stimulating discussions with Jia-Xin Zhang. JX and ZW were supported by MOST of China (Grant No. 2017YFA0302902); ZZ was supported by the National Natural Science Foundation of China (Grant No.12074375), Innovation Program for Quantum Science and Technology (Grant No. 2-6), the Strategic Priority Research Program of CAS (Grant No.XDB33000000) and the Fundamental Research Funds for the Central Universities. 

\bibliography{refs}

\end{document}


\title{Supplemental Material for ``Hubbard Model on Triangular Lattice: Role of Charge Fluctuations''}

\author{Ji-Si Xu}
\affiliation{Institute for Advanced Study, Tsinghua University, Beijing 100084, China}

\author{Zheng Zhu}
\email{zhuzheng@ucas.ac.cn}
\affiliation{Kavli Institute for Theoretical Sciences, University of Chinese Academy of Sciences, Beijing 100190, China}
\affiliation{CAS Center for Excellence in Topological Quantum Computation, University of Chinese Academy of Sciences, Beijing, 100190, China}

\author{Kai Wu}
\affiliation{Manifold Creative Global, 100 Lorong 23 Geylang, 388398, Singapore} 

\author{Zheng-Yu Weng}
\email{weng@mail.tsinghua.edu.cn}
\affiliation{Institute for Advanced Study, Tsinghua University, Beijing 100084, China}

\date{June 19, 2023}

\maketitle

\section{Sign structure in partition function}

In the main text, the partition function of the Hubbard model is generally expressed as 
\begin{gather}
Z_\Hub \equiv \tr(\ee^{- \beta H_\Hub}) = \sum_{c} S[c] W[c]
\tag{\ref{M:eq:ZtoSW}}
\end{gather}
where for any closed path $c$ of the multi-coordinates of spins and chargons, $W[c]$ denotes the positive weight while $S[c]$ collects the total sign.
Formally the sign structure $S[c]$ can be determined by using the high-temperature expansion of the partition function  
\begin{gather}
Z_\Hub \equiv \tr(\ee^{- \beta H_\Hub}) 
= \sum_{n=0}^{\infty} \frac{\beta^n}{n!} \tr((-H_\Hub)^n) 
= \sum_{n=0}^{\infty} \sum_{\set{\alpha_i}_{i=1}^n} 
    \frac{\beta^n}{n!} \prod_{i} \bra{\alpha_{i+1}} (- H_\Hub) \ket{\alpha_i}
\equiv \sum_{c} Z[c]
\end{gather}
with
\begin{gather}\label{ZC}
Z\big[c=\set{\alpha_i}_{i=1}^n\big] \equiv 
    \frac{\beta^n}{n!} \prod_{i} \bra{\alpha_{i+1}} (- H_\Hub) \ket{\alpha_i}
\end{gather}
where each $\ket{\alpha_i}$ runs over a complete set of basis, and $c = \set{\alpha_i}_{i=1}^n$ is a loop in the basis space which will be an Ising spin-chargon representation. Then, $W[c] \equiv  |Z[c]|$ and $S[c] \equiv Z[c]/W[c]$.

First, we point out that $S[c]$ here will only count the signs of all the off-diagonal matrix elements or the hopping term $H_t$ in the Hubbard model, without the contribution from the diagonal elements of the interaction term $H_U$, whose signs can be absorbed by some suitable resummation as shown below.

\subsection{Diagonal (Hubbard interaction) term}

In general, for any Hamiltonian $H$, its partition function $Z$ can be expanded in this form $Z = \sum_{c} S[c] W[c]$, while $S[c]$ collects only the sign of non-diagonal matrix elements of $(-H)$, that is $\bra{\alpha_{i+1}} (- H)\ket{\alpha_i}$ where $\alpha_{i+1} \ne \alpha_{i}$. 
The following is a brief proof. 

First we give a simple physical argument that the sign from the diagonal elements is not important. Here $H$ is Hermitian, and the diagonal elements $\bra{\alpha}H\ket{\alpha}$ must be real. Then, consider adding a constant energy to the Hamiltonian $H^* = H+E_0$, which should not modify the property of the system. 
It keeps the non-diagonal term of $H$, and shifts the diagonal term as $\bra{\alpha}H^*\ket{\alpha} = \bra{\alpha}H\ket{\alpha} + E_0$. 
Now as long as the eigenvalue of $H$ have a upper bound $E_{\mathrm{sup}}$, we can always let $E_0 < -E_{\mathrm{sup}}$, make all diagonal matrix elements $\bra{\alpha}H^*\ket{\alpha}$ negative. 
If we count the sign structure $S[c]$ for $H^*$, the contribution of the diagonal term $\bra{\alpha}(-H^*)\ket{\alpha}$ is always positive and has vanishing sign, while the non-diagonal term remains the same. 
Thus, since the physical property $H^*$ is the same as $H$, the sign of these diagonal terms should be irrelevant. 
This requirement of the existence of energy upper bound is generally satisfied for any finite lattice system, and we may further take it to the thermodynamic limit. 

The following is a formal proof based on the appendix in Ref. \cite{Zhang2014Sign}. 
Here we relax the condition of $c$ to be a closed loop, and let $c=\set{\alpha_i}_{i=0}^n$ where $\alpha_0$ and $\alpha_n$ may not be equal. 
Now for any path $c$, we can split out the stationary steps $\alpha_{i+1} = \alpha_{i} $ and get a squeezed path $\tilde{c} = \set{\tilde{\alpha}_i}_{i=0}^m$. 
$\tilde{c}$ only contains the non-diagonal part, and $\set{\alpha_n, \alpha_{n-1}, \cdots, \alpha_1, \alpha_0} = \set{\tilde{\alpha}_m, \cdots, \tilde{\alpha}_m, \tilde{\alpha}_{m-1}, \cdots, \tilde{\alpha}_{m-1}, \cdots \cdots, \tilde{\alpha}_1, \cdots, \tilde{\alpha}_1, \tilde{\alpha}_0, \cdots, \tilde{\alpha}_0}$.
Now the contribution of path $c$ in the partition function is
\def\ST#1#2{(-H_{\alpha_{#1}\alpha_{#2}})}
\def\SST#1#2{(-H_{\tilde{\alpha}_{#1}\tilde{\alpha}_{#2}})}
\begin{gather}
\begin{split}
Z[c] =& \frac{\beta^n}{n!} \ST{n}{n-1} \cdots \ST{2}{1} \ST{1}{0} \\
    = & \frac{\beta^n}{n!} 
        \SST{m}{m}^{k_{m}} \SST{m}{m-1} \SST{m-1}{m-1}^{k_{m-1}} 
        \cdots \SST{2}{1} \SST{1}{1}^{k_1} 
        \SST{1}{0} \SST{0}{0}^{k_0} \\
    = & \bigg[ \frac{\beta^m}{m!} \SST{m}{m-1} \cdots 
                                    \SST{2}{1} \SST{1}{0} \bigg]
        \bigg[ \frac{\beta^{n-m} m!}{n!}  
            \prod_{i=0}^{m} \SST{i}{i}^{k_i} \bigg] \\
    = & Z[\tilde{c}]
        \bigg[ \frac{m!}{(m+\sum_{i=0}^{m} k_i)!}  
            \prod_{i=0}^{m} 
                (-\beta H_{\tilde{\alpha}_{i}\tilde{\alpha}_{i}})^{k_i} 
        \bigg] 
\end{split}
\end{gather}
where $k_i$ is the number of step stay at $\tilde{\alpha}_i$, which is the number of repetitions of $\tilde{\alpha}_i$ in the path $c$. 
We also denote $H_{\alpha' \alpha}$ as shorthand for $\bra{\alpha' }H\ket{\alpha}$.
Here $Z[\tilde{c}]$ is the contribution of the squeezed path, we should note that it differs from $Z[c]$ by a factor that is only related to the diagonal elements of $H$.

Next we sum all paths $c$ which are squeezed to the same $\tilde{c}$. 
To do this we just sum all possible $\set{k_i}$ configuration where $k_i = 0,1,2\cdots$, and we have
\begin{gather}
\tilde{Z}[\tilde{c}] 
= Z[\tilde{c}] F\big(
    (-\beta H_{\tilde{\alpha}_{0}\tilde{\alpha}_{0}}),
    (-\beta H_{\tilde{\alpha}_{1}\tilde{\alpha}_{1}}),
    \cdots,
    (-\beta H_{\tilde{\alpha}_{m}\tilde{\alpha}_{m}}) \big) \\
F(x_0, x_1, \cdots, x_m)    
= \sum_{k_m} \cdots \sum_{k_0}
    \frac{m!}{(m+\sum_i k_i)!} \prod_{i=0}^{m} x_i^{k_i} \label{Eq:I:F}
\end{gather}
We can prove that the factor $F(x_0, x_1, \cdots, x_m)$ is positive for any real $\set{x_i}$. Inspired by the energy shift argument above, we add a const $a$ to each $x_i$, 
\begin{gather}
\begin{aligned}
& F(x_0+a, x_1+a, \cdots, x_m+a) & \\
= & \sum_{\set{k_i}}
    \frac{m!}{(m+\sum_i k_i)!} \prod_{i=0}^{m} (x_i+a)^{k_i} 
=   \sum_{\set{k_i}}
    \frac{m!}{(m+\sum_i k_i)!} \prod_{i=0}^{m}
    \Bigg[ \sum_{p_m=0}^{k_m} \cdots \sum_{p_0=0}^{k_0} 
            \binom{k_i}{p_i} x_i^{p_i} a^{k_i-p_i} \Bigg] \\
= & \sum_{\set{p_i}} \sum_{\set{q_i}}
    \frac{m!}{(m+\sum_i p_i+\sum_i q_i)!} \prod_{i=0}^{m}
        \binom{p_i+q_i}{p_i} x_i^{p_i} a^{q_i} 
=   \sum_{\set{p_i}} \sum_{Q} \sum_{\sum_i q_i = Q}
    \frac{m!}{(m+\sum_i p_i+Q)!} a^{Q} 
    \prod_{i=0}^{m} \binom{p_i+q_i}{p_i} x_i^{p_i} \\
= & \sum_{\{p_i\}} \sum_{Q} 
    \frac{m!}{(m+\sum_i p_i+Q)!} 
    \Bigg(\prod_{i=0}^{m} x_i^{p_i}\Bigg) a^{Q} 
    \Bigg[ 
        \sum_{\sum_i q_i = Q} \prod_{i=0}^{m} \binom{p_i+q_i}{p_i} \Bigg] 
        &
\end{aligned}
\end{gather}
The last factor in $F$ seems to complicate, but we can prove that it has a very simple form as
\begin{gather}
A_m(p_0, p_1, \cdots, p_m, Q) 
= \sum_{\sum_i q_i = Q} \prod_{i=0}^{m} \binom{p_i+q_i}{p_i} 
= \binom{m+\sum_i p_i+Q}{Q} \label{Eq:I:A}
\end{gather}
To show this, noticing that $A_m$ is totally symmetric to $\set{p_i}$, and the recurrence relation
\begin{gather}
\begin{aligned}
& A_m(p_0, p_1, \cdots, p_m, Q) \\
= & \sum_{\sum_i q_i = Q} 
        \binom{p_0+q_0}{p_0} \prod_{i=1}^{m} \binom{p_i+q_i}{p_i} 
=   \sum_{\sum_i q_i = Q} 
    \Bigg[ \binom{p_0+q_0-1}{p_0-1} + \binom{p_0+q_0-1}{p_0} \bigg]
    \prod_{i=1}^{m} \binom{p_i+q_i}{p_i} \\
= & A_m(p_0-1, p_1, \cdots, p_m, Q) + A_m(p_0, p_1, \cdots, p_m, Q-1)
\end{aligned}
\end{gather}
with the boundary conditions
\begin{gather}
A_m(0, 0, \cdots, 0, Q) 
    = \sum_{\sum_i q_i = Q} \prod_{i=0}^{m} 1 
    = \binom{m+Q}{Q} = \binom{m+0+Q}{Q}  \\
A_m(p_0, p_1, \cdots, p_m, 0) 
    = \prod_{i=0}^{m} \binom{p_i+0}{p_i} = 1 = \binom{m+\sum_i p_i+0}{0} 
\end{gather}
Now we can prove Eq. (\ref{Eq:I:A}) by induction. 
Substitute $A_m$ into $F$ in Eq. (\ref{Eq:I:F}), we have
\begin{gather}
\begin{aligned}
& F(x_0+a, x_1+a, \cdots, x_m+a) 
=   \sum_{\{p_i\}} \sum_{Q} 
    \frac{m!}{(m+\sum_i p_i+Q)!} 
    \Bigg(\prod_{i=0}^{m} x_i^{p_i}\Bigg) a^{Q} 
    \binom{m+\sum_i p_i+Q}{Q} \\
= & \sum_{\{p_i\}} \sum_{Q} 
    \frac{m!}{Q!(m+\sum_i p_i)!} 
    \Bigg(\prod_{i=0}^{m} x_i^{p_i}\Bigg) a^{Q} 
=   \Bigg( \sum_{\{p_i\}} \frac{m!}{(m+\sum_i p_i)!} 
        \prod_{i=0}^{m} x_i^{p_i} \Bigg)
    \Bigg( \sum_{Q} \frac{a^{Q}}{Q!} \Bigg) \\
= & F(x_0, x_1, \cdots, x_m) \ee^{a}
\end{aligned}
\end{gather}
When all $\set{x_i}$ is positive, by definition of $F$, it is positive. For any other $\set{x_i}$, we can choose a large enough $a$, such that $x_i+a>0$ for all $x_i$, and $F(x_0, x_1, \cdots, x_m) = F(x_0+a, x_1+a, \cdots, x_m+a) \ee^{-a} > 0$. 

Now the partition function can be reexpressed as a sum over all squeezed loop $\tilde{c}$ as $Z = \sum_{\tilde{c}} \tilde{Z}[\tilde{c}]$, where $\tilde{Z}[\tilde{c}] = Z[\tilde{c}] F[\beta, \tilde{c}]$ have the same sign factor as $Z[\tilde{c}]$, which only contain the sign of non-diagonal part of $H$. Now the proposition given in the beginning of this subsection is proven.

For the Hubbard model in this work, the interaction term $H_U = U \sum_{i} n_{i\SU}n_{i\SD}$ is diagonal in the Fock basis of electrons, and all the sign we need is in the hopping term $H_t$. 
This property comes with many important results, such as the sign structure we deduced in this work is universal for all interaction strength $U$, and even for other forms of interaction so long as they can be written in the form of particle number operators and is diagonal under the Fock basis.

We stress that although the diagonal terms cannot change the sign structure directly, it does not mean it has no effect on the physical properties. 
By changing the weight of different paths it can dramatically change the system behavior and may alter the interpretation of the sign structure. 
For example, in the large $U/t$ limit of the Hubbard model, the interaction term suppresses the weight of paths that have long-range charge fluctuation, which effectively suppresses the fermionic sign of the electrons. 
This is exactly how the Heisenberg model as a large-$U$ Hubbard model at half-filling becomes sign free (except for a geometric phase for a non-bipartite lattice, see below) and how phase-string effect emerges upon doping or reducing $U$. 

\subsection{Sign structure of the AFM Heisenberg model}

Our discussion starts from the sign structure of the large-$U$ Hubbard model, which reduces to the AFM Heisenberg model and serves as a reference limit for the sign structure. 

Introducing the Schwinger-boson representation $\mathbf{S} = \frac{1}{2} \bd_{\alpha} \boldsymbol{\sigma}_{\alpha\beta}\bo_{\beta}$ to the Heisenberg model $H_J$, one has

\begin{gather}
H_J 
\equiv J \sum_{\nn{ij}}
    (\mathbf{S}_{i} \cdot \mathbf{S}_{j} - \frac{1}{4})
= - \frac{J}{2} \sum_{\ave{ij}} 
       [(-\bd_{i\SU}\bo_{j\SU}) \bd_{j\SD}\bo_{i\SD} 
        + \bd_{i\SD}\bo_{j\SD} (-\bd_{j\SU}\bo_{i\SU})] + (\DT)
\end{gather}
where $\DT$ denotes the diagonal term, whose contribution to the sign structure can be ignored as discussed above. 
The rest two off-diagonal terms describe the swap process between an up-spinon and a down-spinon. 
Note that in the matrix elements of $(-H_J)$, a minus sign can be associated with the hopping of up-spinons $\bd_{i\SU}\bo_{j\SU}$, which contributes a minus sign to the sign structure $S[c]$ in the partition function according to the high-temperature expansion outlined above. It is easy to verify that the total sign structure $S[c]$ in this Heisenberg limit is equal to $S_0[c]$ as follows
\begin{gather}
S[c] = (-1)^{N^\SU[c]}
\end{gather}
where $N^\SU[c]$ is the number of steps of the $\SU$-spinon hopping in a closed path $c$.

One may also rewrite $(- \bd_{i\SU}\bo_{j\SU})$ as $(\ee^{\iu \phi^\gamma_{ij}} \bd_{i\SU}\bo_{j\SU})$ and let $\phi^\gamma_{ij} = \pm \pi$.
Then the sign structure can be interpreted as a geometric phase of an up-spinon moving on a fictitious background field $\phi^\gamma_{ij}$. 
For a bipartite lattice like a square lattice, each plaquette consists of an even number of bonds, and the flux for the plaquette must be an integer multiple of $2\pi$, which is trivial. It means the AFM Heisenberg model on a bipartite lattice is essentially sign-free, consistent with the well-known Marshall theorem \cite{Marshall1955Antiferromagnetism}.
For a non-bipartite lattice, the geometric phase remains non-trivial. In particular, in a triangular lattice, each unit triangle contains a $\pi$ flux. This geometric phase is very important to the physics of triangular lattice, including the formation of the 120\degree\ AFM order. 

\subsection{Sign structure due to the hopping term at arbitrary $U$}

The sign structure $S[c]$ for the Hubbard model at a finite $U$ will then only depend on the hopping term $H_t$, independent of the Hubbard interaction term $H_U$ as shown above. In the following, we will present the derivation of the sign structure for the Hubbard model as given in Eq. (\ref{M:eq:SPStri}) of the main text. Note that the sign structure for the Hubbard model on a square lattice has been previously determined in Ref. \onlinecite{Zhang2014Sign}. Below a generalized version that works for any lattices including the triangular lattice will be obtained.

Starting with the usual slave-fermion decomposition of the electron $c$-operator
\begin{gather}
\cd_{i\sigma} = \bd_{i\sigma}\ha_i + \sigma \dd_i\bo_{i\bar\sigma} 
\label{eq:frac}
\end{gather}
where $\bar\sigma = -\sigma$, and $\bd_{i\sigma}$ denotes a bosonic spinon (Schwinger boson) creation operator, while $\dd_i$ and $\hd_i$ are fermionic chargon (doublon and holon) operators at site $i$. At each site they obey the single occupancy constrain $\bd_{i\SU}\bo_{i\SU}+\bd_{i\SD}\bo_{i\SD}+\dd_{i}\da_{i}+\hd_{i}\ha_{i} = 1$. Then one may express the hopping Hamiltonian $H_t$ in a fractionalized form 
\begin{gather}
\begin{aligned}
H_t 
&=  - t \sum_{\ket{j,i},\sigma} 
        \cd_{i\sigma}\co_{j\sigma} \\
& =  - t \sum_{\ket{j,i},\sigma} 
        (\bd_{i\sigma}\ha_i + \sigma \dd_i\bo_{i\bar\sigma})
        (\hd_j\bo_{j\sigma} + \sigma \bd_{j\bar\sigma}\da_j) \\
& \equiv T + P 
\end{aligned} \label{eq:HtTP}
\end{gather}
where 
\begin{align}
T \equiv 
& - t \sum_{\ket{j,i},\sigma} 
    \bd_{i\sigma}\bo_{j\sigma}( - \hd_j\ha_i + \dd_j\da_i ) \\
P \equiv 
& - t \sum_{\ket{j,i},\sigma} 
    \sigma \bd_{i\sigma}\bd_{j\bar\sigma} \ha_i\da_j + \HC 
\end{align}

Let us first examine the contribution of the hopping matrix $\bra{\alpha_{i+1}} (- P) \ket{\alpha_i}$ in Eq. (\ref{ZC}). One has
\begin{gather}
\begin{aligned}
P 
& = - t \sum_{\ket{j,i}} (
    - \bd_{i\SU}\bd_{j\SD} \da_j\ha_i)
    + (\bd_{i\SD}\bd_{j\SU} \da_j\ha_i )+ \HC 
\end{aligned}    
\end{gather}
Thus a process of creation of an $\SU$-spinon at $i$ and annihilation of a doublon at the nearest-neighbor site $j$ will contribute to an additional sign of $(-1)$ as if $(-1)$ is always associated with an $\SU$-spinon which hops from the doublon site to the holon site. Note that in the large-$U$ limit, $P$ will be the sole leading hopping term at half-filling to result in the superexchange term of the Heisenberg model, in which the above sign $(-1)$ can be indeed counted by $S_0[c]=(-)^{N^\SU[c]}$ depending on the hopping steps of the up-spin for a given closed loop $c$ as shown in the previous subsection.  

Then we may rewrite the $T$ term by
\begin{gather}
\begin{aligned}\label{T}
T =
&  - t \sum_{\ket{j,i},\sigma} 
    (-\sigma \bd_{i\sigma}\bo_{j\sigma}) \sigma ( \hd_j\ha_i - \dd_j\da_i) 
\end{aligned}    
\end{gather}
and examine the hopping matrix $\bra{\alpha_{i+1}} (- T) \ket{\alpha_i}$ in Eq. (\ref{ZC}). Note that in $T$ [Eq. (\ref{T})], $(-\sigma \bd_{i\sigma}\bo_{j\sigma})$ describes the hopping of bosonic spinon with a $(-1)$ sign for each up-spinon hopping in consistency with the counting in the process $P$ above, giving rise to a full $S_0[c]$ as the geometric phase involving the hopping of the spinon in $H_t$. Then there is always an additional sign $\sigma=\pm 1$, which should be counted as associated with the hopping of the holon, or a sign $-\sigma$ associated with the doublon in the hopping $T$ in Eq. (\ref{T}). 
In other words, the contribution of the hopping $T$ to the sign structure, besides the geometric phase of the spinon in $S_0[c]$, will be the phase-string sign factor: $\tau_{\mathrm{ph}}[c] = (-1)^{N_{\SD}^{h}[c]} \times (-1)^{N_{\SU}^{d}[c]}$. And finally, one should notice that since the slave-fermion representation is used in the above expansion, one needs to also count the fermionic sign of chargon $h$ and $d$, which may be written as $(-)^{N_{\mathrm{ex}}^{\mathrm{ch}}[c]}$, where $N_{\mathrm{ex}}^{\mathrm{ch}}[c]$ counts the swaps between identical chargon operators (holons and doublons, respectively).

Therefore, the total sign structure in the partition function of the Hubbard model is composed of three parts as given in Eq. (\ref{M:eq:SPStri}) of the main text:
\begin{gather}
S[c] \equiv S_0[c] \times (-1)^{N_{\mathrm{ex}}^{\mathrm{ch}}[c]} \times \tau_{\mathrm{ph}}[c] 
\tag{\ref{M:eq:SPStri}}
\end{gather}

It is noted that the geometric phase $S_0[c] = (-)^{N^\SU[c]}$ obtained above counts the minus signs for two types of processes, i.e.,
the hopping of an $\SU$-spinon by exchanging with a chargon (via $T$), and an $\SU$-spinon hopping from a doublon site to a holon site via $P$ or \emph{vice versa}, which correspond to a free hopping of an $\SU$-spinon irrespective of single or double occupancy.  
The associated sign structure $S_0[c]$ is thus regarded as a pure geometric phase for the $\SU$-spinon as it does not differentiate the ``backflow'' during its hopping. In the large $U/t$ limit, with the suppression of the charge fluctuation, such a geometric phase is the only remaining sign structure for the triangular lattice case. On the other hand, for the bipartite lattice, the geometric phase must vanish at half-filling as discussed in the previous subsection with $S_0[c] = 1$ (Ref.\cite{Zhang2014Sign}). 

\section{The \texorpdfstring{$\sigma$}{sigma}-hopping term}

In the main text, we demonstrate that the phase-string $\tau_{\mathrm{ph}}$ in $S[c]$ will play a critical role to mediate mutual entanglement between the spin and charge degrees of freedom in the Hubbard model with the decrease of $U$. For this purpose, a so-called $\sigma$-Hubbard model, $H_{\sigma-\Hub}$, is constructed, in which the phase-string $\tau_{\mathrm{ph}}$ is precisely removed from the sign structure without changing the rest of the sign structure as well as the positive weight $W[c]$ for the Hubbard model. This method has been applied to the square lattice Hubbard model\cite{He2016Possibility}. A straightforward proof is given as follows.

In the Hubbard model, the hopping term $H_t$ can be split into two parts, $T$ and $P$, as shown above, in which the phase-string is only related to the single chargon hopping term $T$. In the $\sigma$-Hubbard model, a sign change inside the hopping term $H_{\sigma t}$ can solely cancel out such a phase-string effect.

For the hopping term, one has 
\begin{gather}
\begin{aligned}
\cd_{i\sigma}\co_{j\sigma} 
& = (\bd_{i\sigma}\ha_i + \sigma \dd_i\bo_{i\bar\sigma})
    (\hd_j\bo_{j\sigma} + \sigma \bd_{j\bar\sigma}\da_j) \\
& = \bd_{i\sigma}\bo_{j\sigma} \ha_i\hd_j
    + \bo_{i\bar\sigma}\bd_{j\bar\sigma} \dd_i\da_j
    + \sigma \bd_{i\sigma}\bd_{j\bar\sigma} \ha_i\da_j
    + \sigma \bo_{i\bar\sigma}\bo_{j\sigma} \dd_i\hd_j 
\end{aligned} \label{eq:Ht4T}
\end{gather}
Here it is split into four terms. To cancel the phase-string one only needs to add a minus sign to $\bd_{i\SD}\bo_{j\SD} \ha_i\hd_j$ and $\bo_{i\SU}\bd_{j\SU} \dd_i\da_j$. 
Or equivalently, a $\sigma$ sign is added before $\bd_{i\sigma}\bo_{j\sigma} \ha_i\hd_j + \bo_{i\bar\sigma}\bd_{j\bar\sigma} \dd_i\da_j$, which is the $T$ part of the hopping term. 

In Eq. (\ref{M:eq:Hst}) of main text, an projection operator $\hat{P}^{T}_{ij}$ is introduced. Define $\hat{P}^{T}_{ij}$ as 
\begin{gather}
\hat{P}^{T}_{ij} = 
    (m_{i\SU} m_{i\SD}) (m_{j\SU} n_{j\SD})
    + (m_{i\SU} m_{i\SD}) (n_{j\SU} m_{j\SD})
    + (m_{i\SU} n_{i\SD}) (n_{j\SU} n_{j\SD})
    + (n_{i\SU} m_{i\SD}) (n_{j\SU} n_{j\SD})
\end{gather}
where $n_{i\sigma} = \cd_{i\sigma}\co_{i\sigma}$ and $m_{i\sigma} = 1-n_{i\sigma}$. 
Then the four terms in $\hat{P}^{T}_{ij}$ are projected to the four states $\ket{\CH}_i\ket{\SD}_j$, $\ket{\CH}_i\ket{\SU}_j$, $\ket{\SD}_i\ket{\CD}_j$, $\ket{\SU}_i\ket{\CD}_j$, respectively. 
Apply $\cd_{i\sigma}\co_{j\sigma}$ to these two state is equivalent to $T$ part of hopping term. Thus we can define the $\sigma$-hopping term $H_{\sigma t}$ as in Eq. (\ref{M:eq:Hst}):
\begin{gather}
\cd_{i\sigma}\co_{j\sigma} \to
    \cd_{i\sigma}\co_{j\sigma} 
        [\sigma \hat{P}^{T}_{ij}+(1-\hat{P}^{T}_{ij})]
\end{gather}

Another equivalent way is splitting the hopping term as
\begin{gather}
\cd_{i\sigma}\co_{j\sigma} = 
  \big( m_{i\bar\sigma} \cd_{i\sigma}\co_{j\sigma} m_{j\bar\sigma} \big)
+ \big( n_{i\bar\sigma} \cd_{i\sigma}\co_{j\sigma} n_{j\bar\sigma} \big)
+ \big( m_{i\bar\sigma} \cd_{i\sigma}\co_{j\sigma} n_{j\bar\sigma} \big)
+ \big( n_{i\bar\sigma} \cd_{i\sigma}\co_{j\sigma} m_{j\bar\sigma} \big)
\end{gather}
It can be shown that these four terms correspond four terms in Eq. (\ref{eq:Ht4T}). Now we can cancel the phase-string sign factor as
\begin{gather}
\cd_{i\sigma}\co_{j\sigma} \to
\sigma 
  \big(m_{i\bar\sigma} \cd_{i\sigma}\co_{j\sigma} m_{j\bar\sigma} \big)
+ \sigma
  \big(n_{i\bar\sigma} \cd_{i\sigma}\co_{j\sigma} n_{j\bar\sigma} \big)
+ \big(m_{i\bar\sigma} \cd_{i\sigma}\co_{j\sigma} n_{j\bar\sigma} \big)
+ \big(n_{i\bar\sigma} \cd_{i\sigma}\co_{j\sigma} m_{j\bar\sigma} \big)
\end{gather}
This form is used in the implementation of the numerical calculation.  

\section{About numerical calculation}

As mentioned in the main text, we utilize a numerical method, DMRG specifically, to verify our argument. 
This section will provide the necessary technical details for the interpretation of these numerical results.

\begin{figure}[ht]
    \centering
    \includegraphics[width=3.4in]{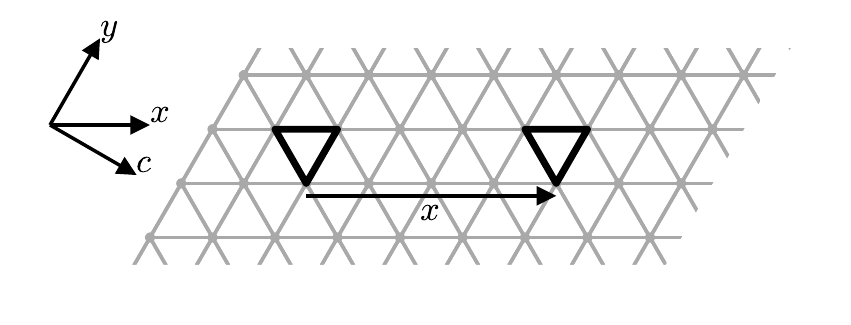}
    \caption{The lattice geometry used in numerical computation. 
    We choose open boundary condition in $x$ direction and periodical boundary condition in $y$ direction. 
    The lattice forms a cylinder, and its axial direction is indicated as $c$. 
    The circumference around $y$ direction $L_y$ is fixed as $4$. 
    The two black triangles illustrate the choice of triangle configuration when measuring the chiral order correlation function. }
    \label{fig:lattice}
\end{figure}

\paragraph*{Model and DMRG setup.} 
We choose the lattice geometry as a four-leg ladder, as shown in Fig. \ref{fig:lattice}. 
The lattice is open along $x$ direction with length $L_x$, and closed in $y$ direction with circumference $L_y=4$. 
During the DMRG calculation, we set the tunable bond dimension $\mathcal{D}$, which increases with DMRG sweeps and the maximal $D=24000$ for the Hubbard model and  $\mathcal{D}=12000$ for $\sigma$-Hubbard model, and the maximal accumulated truncation error is less than $10^{-7}$.
With all this evidence together, we believe the numerical results are converged and reliable.  

\newcommand{\vS}{\mathbf{S}}

\begin{figure}[ht]
    \centering
    \begin{minipage}[t]{0.49\textwidth}
        \raggedright (a) \\
        \centering
        \includegraphics[width=3.4in]{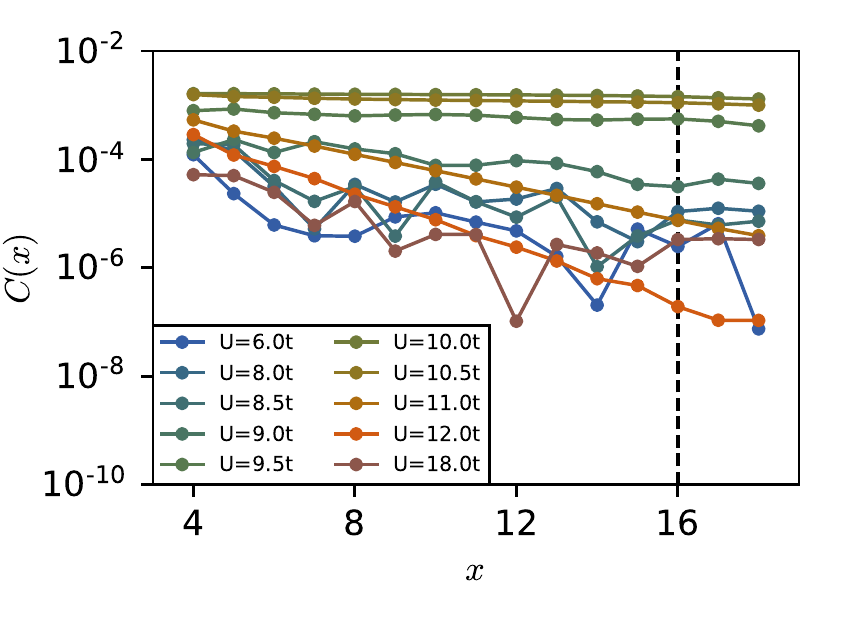}        
    \end{minipage}
    \begin{minipage}[t]{0.49\textwidth}
        \raggedright (b) \\
        \centering
        \includegraphics[width=3.4in]{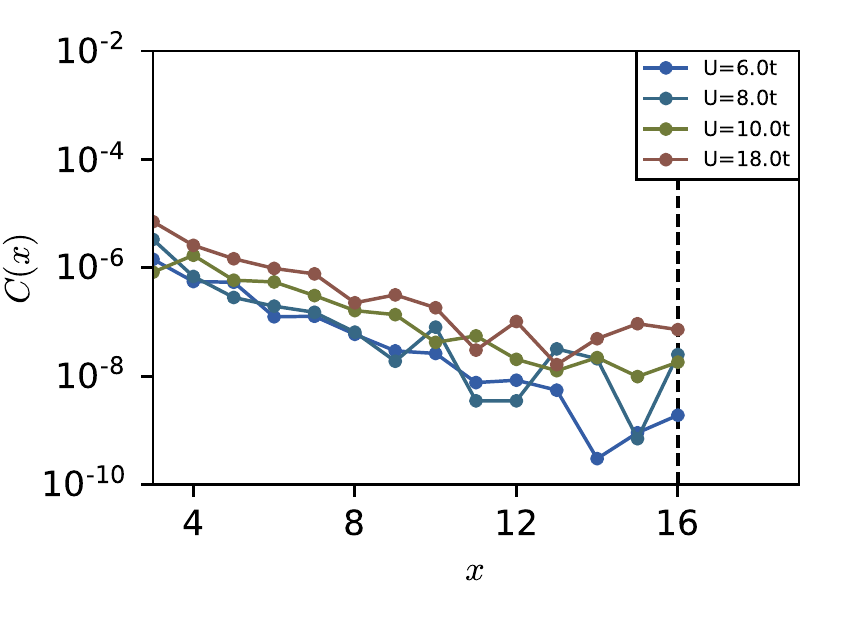}
    \end{minipage}
    \caption{Correlation function of the chiral order parameter of (a) Hubbard model and (b) $\sigma$-Hubbard model. 
    The vertical line in the background indicates $x_R$, which is the reference point to estimate the absolute value of chiral order parameter $|\ave{\chi(i)}|$.}
    \label{fig:chiral}
\end{figure}

\paragraph*{Chiral order parameter.}
In CSL phase, the chiral spin order parameter $\chi(i) = \vS_{i} \cdot (\vS_{i+\hat{y}} \times \vS_{i+\hat{y}-\hat{x}})$ will have nonzero average due to symmetry breaking.
In our case, the finite size effect will suppress the symmetry breaking, and $\ave{\chi(i)}$ is always zero. 
As a roundabout, we calculate the correlation function $C(x) = \ave{\chi(i) \chi(i+x)}$ first. 
In Fig. \ref{fig:lattice}, we show the spatial configuration sites we used in the definition of $\chi(i)$ and $C(x)$.
When entering CSL phase, $C(x)$ can sustain a finite value over a long distance, and otherwise decay exponentially, as shown in Fig. \ref{fig:chiral}. 
We choose a reference point $x_R = 16$, and estimate the order parameter as $|\ave{\chi(i)}| = \sqrt{C(x_R)}$. The results are shown in Figs. \ref{M:fig:spin}(b) and (c) of the main text.

\begin{figure}[ht]
    \centering
    \begin{minipage}[t]{0.49\textwidth}
        \raggedright (a) \\
        \centering
        \includegraphics[width=3.4in]{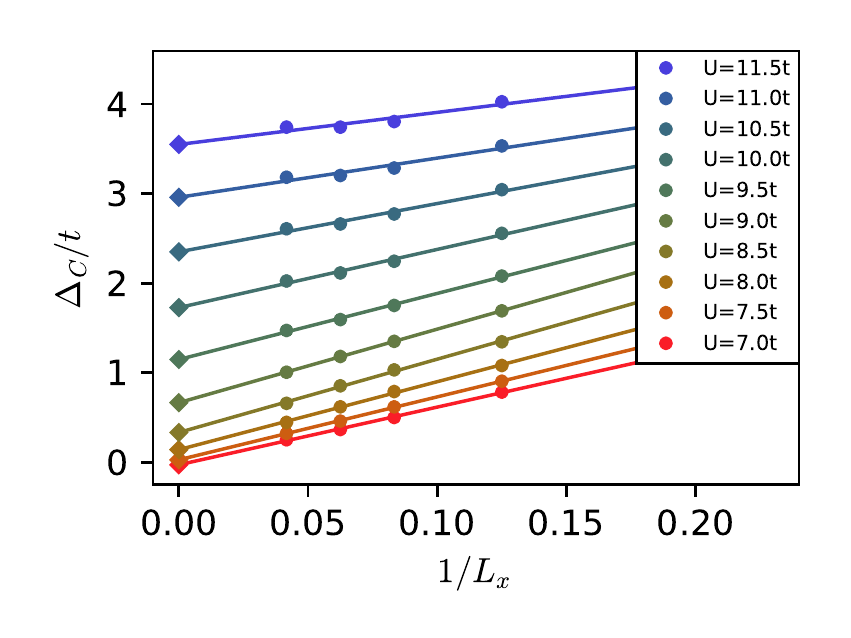}        
    \end{minipage}
    \begin{minipage}[t]{0.49\textwidth}
        \raggedright (b) \\
        \centering
        \includegraphics[width=3.4in]{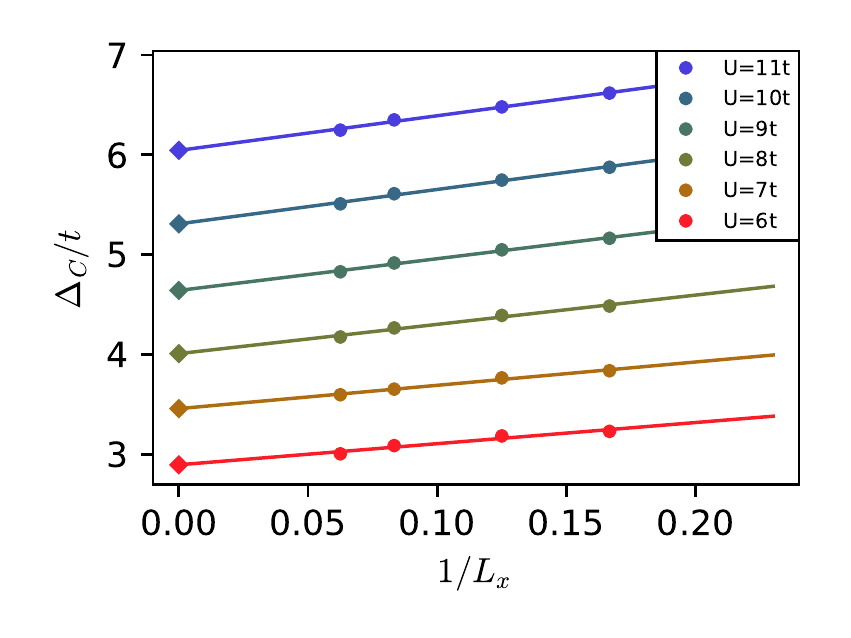}
    \end{minipage}
    \caption{Figures of charge gap to lattice length for (a) Hubbard model and (b) $\sigma$-Hubbard model. 
    The round points mark the results obtained from DMRG calculation, the lines are the linear fitting to $1/L_x$.
    The diamond points on the left indicate the extrapolated charge gap from the fitting.}
    \label{fig:gap}
\end{figure}

\paragraph*{Charge gap.} 
We measure the charge gap of the model by calculating the energy difference of the ground state when adding or removing electrons, as $\Delta_\TxC = \frac{1}{2}[E_0(N+1,N+1) + E_0(N-1,N-1) - 2 E_0(N,N)]$. 
Here, $E_0(N_\SU, N_\SD)$ is the ground state energy with $N_\SU$ spin-up electrons and $N_\SD$ spin-down electrons, and $N$ is the lattice site number. 
$E_0(N,N)$ is the ground state energy of half-filling. 
Note that we add or remove a pair of electrons with opposite spins to cancel the possible effect from spin gap. 
The finite size effect will lead to the overestimation of the charge gap, and to overcome this we apply the standard finite size scaling procedure. 
As shown in Fig. \ref{fig:gap}, we calculate the $\Delta_\TxC$ for different lattice lengths $L_x$, and fit them linearly to $1/L_x$ as $\Delta_\TxC(L_x) = \Delta_\TxC^\infty + b /L_x$, and $\Delta_\TxC^\infty$ is the estimation for the charge gap under thermodynamic limit as we extrapolate to $L_x$ to infinity. 
For the Hubbard model, the charge gap approach to zero after the extrapolation when in the metallic phase, which is consistent with the Fermi liquid behavior. 
The summarized results are shown in Fig. \ref{M:fig:charge}(a) of the main text.
Fitting with a quadratic function of $1/L_x$ gives similar results, except around the $U/t = 8.5$ for the Hubbard model, which is close to the phase boundary.

\paragraph*{Double occupancy.} 
The number of double occupancy site is defined as $\hat{D} = \sum_{i} n_{i\SU}n_{i\SD}$. It has a convenient property that the interaction term of Hubbard model $H_U = U \hat{D}$. Thus we can get its average as $D = \ave{\hat{D}} = \frac{\partial}{\partial U} E_0(U)$, where $E_0(U)$ is the ground state energy with interaction strength $U$. This gives an efficient way to calculate $D$. The results are shown in Fig. \ref{M:fig:charge}(b). 
 
\bibliography{refs}